\documentclass[extra,referee]{gji}
\usepackage{timet}
\usepackage{amsmath}
\usepackage{amsfonts}
\usepackage{upgreek}
\usepackage{graphicx}
\graphicspath{{figures/}} 
\usepackage{setspace}

\usepackage{multirow}
\usepackage[T1]{fontenc}

\usepackage[]{siunitx}
\usepackage{arydshln}
\renewcommand{\arraystretch}{1.7}

\begin{document}

\begin{titlepage}
    \begin{center}
        \vspace*{-2.0cm} 
        

		 \vspace{5cm}
		 
        \huge
        \textbf{Bayesian Variational Time-lapse Full-waveform Inversion}
        
        \vspace{4.0cm}
        \LARGE
        Xin Zhang$^{1,2}$, Andrew Curtis$^{2}$\\

        \vspace{1.0cm}
        \Large
        $^1$School of Engineering and Technology, China University of Geosciences, Beijing, China \\
        $^2$School of GeoSciences, University of Edinburgh, United Kingdom

        \vspace{1cm}
        \Large
        E-mail: \textit{x.zhang2@ed.ac.uk, andrew.curtis@ed.ac.uk}

        \vfill
        \vfill
    \end{center}
    
\end{titlepage}

\newpage

\begin{summary}
Time-lapse seismic full-waveform inversion (FWI) provides estimates of dynamic changes in the subsurface by performing multiple seismic surveys at different times. Since FWI problems are highly non-linear and non-unique, it is important to quantify uncertainties in such estimates to allow robust decision making. Markov chain Monte Carlo (McMC) methods have been used for this purpose, but due to their high computational cost, those studies often require an accurate baseline model and estimates of the locations of potential velocity changes, and neglect uncertainty in the baseline velocity model. Such detailed and accurate prior information is not always available in practice.

In this study we use an efficient optimization method called stochastic Stein variational gradient descent (sSVGD) to solve time-lapse FWI problems without assuming such prior knowledge, and to estimate uncertainty both in the baseline velocity model and the velocity change. We test two Bayesian strategies: separate Bayesian inversions for each seismic survey, and a single join inversion for baseline and repeat surveys, and compare the methods with the standard linearised double difference inversion. The results demonstrate that all three methods can produce accurate velocity change estimates in the case of having fixed (exactly repeatable) acquisition geometries, but that the two Bayesian methods generate more accurate results when the acquisition geometry changes between surveys. Furthermore the joint inversion provides the most accurate velocity change and uncertainty estimates in all cases. We therefore conclude that Bayesian time-lapse inversion, especially adopting a joint inversion strategy, may be useful to image and monitor the subsurface changes, in particular where uncertainty in the results might lead to significantly different decisions.
\end{summary}

\section{Introduction}
A wide variety of academic and practical applications require that we detect property changes in the subsurface in order to understand dynamic processes in the Earth's interior. Time-lapse seismic monitoring provides an important tool for this purpose. This involves conducting multiple seismic surveys acquired at the same site at different times \citep{lumley2001time}. Changes in certain subsurface properties are estimated by computing the difference between models constructed for surveys at different times (the first survey being called the baseline survey and subsequent surveys being called monitoring surveys). In order to assess reliability of the property changes and interpret the results with appropriate levels of confidence, it is also necessary to quantify the uncertainty in such estimates.

Seismic full waveform inversion (FWI) is a method which uses full seismic recordings to characterize properties of the Earth's interior \citep{tarantola1984inversion, tarantola1988theoretical, pratt1999seismic,tromp2005seismic, plessix2006review}, and has been applied at industrial scale \citep{virieux2009overview, prieux2013multiparameter}, regional scale \citep{tape2009adjoint, fichtner2009full} and global scale \citep{french2014whole, lei2020global}. Because of its high spatial resolution, the method has been extended to time-lapse studies to image changes in the subsurface. For example, a range of different schemes have been proposed for this purpose, such as parallel FWI \citep{plessix2010some}, sequential FWI \citep{asnaashari2015time}, double difference FWI \citep{watanabe2004differential, zheng2011strategies} and joint FWI \citep{maharramov2014joint, yang2014time}. However, all of these studies used linearised methods to solve their respective inverse problems and therefore cannot provide accurate uncertainty estimates. As a result, it becomes difficult to interpret the estimated property changes, and to use those estimates for subsequent applications.

Bayesian inference methods provide a variety of different ways to solve inverse problems and can produce accurate uncertainty estimates. In Bayesian inference, prior information is represented by a probability density function (pdf) called the \textit{prior} pdf, which describes information about the parameters of interest prior to conducting the inversion. Bayes' theorem updates the prior pdf with new information contained in the data to construct a so-called \textit{posterior} pdf which describes the total state of information about the parameters post inversion. The updating process is referred to as Bayesian inference.  

Markov chain Monte Carlo (McMC) is one method that is commonly used to solve Bayesian inference problems. The method generates a set (chain) of successive samples from the posterior probability distribution by taking a structured random-walk through parameter space \citep{brooks2011handbook}. Those samples can thereafter be used to calculate statistics of the posterior pdf, for example the mean and standard deviation. The Metropolis-Hastings algorithm is one such method \citep{metropolis1949monte, hastings1970monte, green1995reversible} and has been applied to a range of geophysical applications including gravity inversion \citep{mosegaard1995monte, bosch2006joint}, vertical seismic profile inversion \citep{malinverno2000monte}, electrical resistivity inversion \citep{malinverno2002parsimonious, galetti2018transdimensional}, electromagnetic inversion \citep{minsley2011trans, ray2013robust, blatter2019bayesian}, surface wave dispersion inversion \citep{bodin2012transdimensional, shen2012joint, young2013transdimensional, galetti2017transdimensional, zhang20183}, travel time tomography \citep{bodin2009seismic, galetti2015uncertainty, zhang2020imaging} and full-waveform inversion \citep{ray2017low, sen2017transdimensional, guo2020bayesian}. However, due to its random-walk behavior, the method becomes inefficient in high dimensional parameter spaces (e.g., >1,000 dimensions are commonly encountered in geophysical imaging problems). To reduce this issue, a variety of more advanced methods have been introduced to geophysics, such as Hamiltonian Monte Carlo \citep{duane1987hybrid, fichtner2018hamiltonian, gebraad2020bayesian}, Langevin Monte Carlo \citep{roberts1996exponential, siahkoohi2020uncertainty}, stochastic Newton McMC \citep{martin2012stochastic, zhao2019gradient} and parallel tempering \citep{hukushima1996exchange, dosso2012parallel, sambridge2013parallel}. These allow convergence to be accelerated by assuming specific information about the structure of the problem to be solved. Nevertheless, all of these methods still incur high computational costs and are therefore difficult to use in time-lapse full-waveform inversion. To enable Bayesian inference in time-lapse studies, \cite{kotsi2020uncertainty} exploited a fast, local solver together with the discrete cosine transform to solve time-lapse FWI problems using the Metropolis-Hastings algorithm, and directly imaged the velocity change by assuming a known baseline model. To further improve the efficiency, Hamiltonian Monte Carlo was used to solve the problem but with a regular grid parameterisation \citep{kotsi2020time}. However, these methods require a known baseline model and prior knowledge about the location of potential velocity change zones, which are not always available in practice and which therefore restricts their practical applications.

Variational inference solves Bayesian inference problems in a different way: the method seeks an optimal approximation to the posterior pdf within a predefined (simplified) family of probability distributions, by minimizing the difference between the approximating pdf and the posterior pdf \citep{bishop2006pattern, blei2017variational}. A typical metric used to measure this difference is the Kullback-Leibler (KL) divergence \citep{kullback1951information}; since this metric is minimized the method solves Bayesian inference problems using optimisation, rather than stochastic sampling as used in McMC methods. Consequently for some classes of problems variational inference can be computationally more efficient and provide better scaling to higher dimensional problems \citep{blei2017variational, zhang2018advances}. The method also allows us to take advantage of  stochastic and distributed optimisation \citep{robbins1951stochastic, kubrusly1973stochastic} by dividing large data sets into small minibatches. In addition, variational inference can often be parallelized at the individual sample level, which makes the method even more efficient in real time. By contrast, in McMC one cannot use small minibatches as they break the detailed balance property required by most McMC methods \citep{o2004kendall}, and McMC does not allow parallelization at the sample level as each sample in McMC depends on the previous sample. 

Variational inference has been applied to a range of geophysical applications. \cite{nawaz2018variational} introduced a mean-field variational inference method to invert for geological facies distributions using seismic data \citep{nawaz2019rapid, nawaz2020variational}. While the method is computationally extremely efficient, it neglects correlation information between parameters, and the approach taken in these papers requires bespoke mathematical derivations which restricts its application to the range of problems for which such derivations are possible. To extend variational inference to a wider class of inverse problems, a set of more general algorithms have been proposed. Based on a Gaussian variational family, \cite{kucukelbir2017automatic} proposed a method called automatic differential variational inference (ADVI) which can be applied easily to many inverse problems and has been used to solve travel time tomography \citep{zhang2020seismic} and earthquake slip inversion problems \citep{zhang2022bayesian}. \cite{rezende2015variational} proposed a method called normalizing flows in which one optimizes a sequence of invertible transforms that transform a simple initial probability distribution to any other distribution, which can be used to approximate the posterior probability distribution. In geophysics the method has been applied to travel time tomography \citep{zhao2021bayesian} and image denoising \citep{siahkoohi2020faster}. By using a set of samples (called particles) to represent the approximating distribution, \cite{liu2016stein} proposed the Stein varational gradient descent (SVGD) method which iteratively moves these particles through parameter space so as to minimize the KL divergence, such that in the final state their density approximates the posterior distribution. In geophysics this method has been applied to travel time tomography \citep{zhang2020seismic}, earthquake location inversion \citep{smith2022hyposvi}, hydrogeological inversion \citep{ramgraber2021non} and full waveform inversion \citep{zhang2020variational, zhang2021bayesiana}. More recently, \cite{zhang20233} introduced a variant of the SVGD method called stochastic SVGD (sSVGD) to solve 3D FWI problems and demonstrated that the method can provide more accurate results than ADVI and the original SVGD.  

Based on the results of these studies we chose to test the sSVGD method for the solution of time-lapse FWI problems. In particular, we do not assume prior knowledge about an accurate baseline model, nor about locations of potential velocity change zones, and we estimate uncertainty for both the baseline velocity and the time-lapse velocity change. To solve time-lapse FWI problems, we consider two Bayesian strategies, namely separate Bayesian inversion for baseline and monitor surveys, and joint Bayesian inversion for both surveys together, and compare the results with those from standard double difference inversion. In addition, we perform studies with both fixed and perturbed acquisition geometries between the baseline and monitoring surveys to test the robustness of each method to typical variations in survey design that may occur in practical applications. 

In the following section we first describe the two Bayesian inversion strategies and double difference inversion. In section 3 we apply the suit of methods to a time-lapse FWI problem and compare the results and their computational costs. We use the results to demonstrate that sSVGD can be used to solve Bayesian time-lapse FWI problems and produce accurate velocity change estimates as well as associated uncertainties. We conclude by defining particular contexts in which Bayesian time-lapse FWI provides an important tool to image and monitor subsurface property changes.

\section{Methods}

\subsection{Standard time-lapse full waveform inversion}
The standard way to perform FWI is to minimize a misfit function between observed data $\mathbf{d}$ and model predicted data $\mathbf{u}(\mathbf{m})$ plus a regularization term:
\begin{equation}
L(\mathbf{m}) = \frac{1}{2}|\mathbf{d} - \mathbf{u}(\mathbf{m})|^{2} + \lambda |\mathbf{D}\mathbf{m}|^{2}
\label{eq:fwi_loss}
\end{equation}
where $\mathbf{m} \in \mathbb{R}^{r}$ is the parameterized earth model, $\mathbf{D}$ typically represents a finite-difference derivative matrix and $\lambda$ controls the magnitude of regularization. The most straightforward implementation of time-lapse FWI is to perform the above minimization for each dataset from individual surveys; differences between the obtained models are regarded as estimates of the time-lapse change. In this mode of implementation either both inversions can be performed using the same starting model, or the model obtained from inversion of the baseline survey can be used as the starting model for the monitoring survey inversion. 

A more efficient method is so-called double difference FWI which uses differential data between the two sets of data obtained in the baseline and monitoring surveys \citep{watanabe2004differential, denli2009double, zheng2011strategies}. The misfit function for double difference FWI is:
\begin{equation}
L(\mathbf{m}_{2}) = \frac{1}{2}|(\mathbf{d}_{2} - \mathbf{d}_{1})-\left( \mathbf{u}(\mathbf{m}_{2})-\mathbf{u}(\mathbf{m}_{1}) \right)|^{2} + \lambda |\mathbf{D}\mathbf{m}_{2}|^{2} + \mu |\mathbf{m}_{2}-\mathbf{m}_{1}|^{2}
\label{eq:ddfwi1}
\end{equation}
where we used subscript 1 and 2 to denote variables of the baseline inversion and monitoring inversion respectively, $\lambda$ and $\mu$ are parameters that control the strength of regularization of model $\mathbf{m}_{2}$ and of the difference between $\mathbf{m}_{2}$ and $\mathbf{m}_{1}$, respectively. The above equation can be reformulated as:
\begin{equation}
\begin{aligned}
	L(\mathbf{m}_{2}) &= \frac{1}{2}|(\mathbf{d}_{2} - \mathbf{d}_{1} + \mathbf{u}(\mathbf{m}_{1}))- \mathbf{u}(\mathbf{m}_{2}) |^{2} + \lambda |\mathbf{D}\mathbf{m}_{2}|^{2} + \mu |\mathbf{m}_{2}-\mathbf{m}_{1}|^{2} \\
	&= \frac{1}{2}|\mathbf{d}_{2}^{'}- \mathbf{u}(\mathbf{m}_{2}) |^{2} + \lambda |\mathbf{D}\mathbf{m}_{2}|^{2} + \mu |\mathbf{m}_{2}-\mathbf{m}_{1}|^{2}
	\label{eq:ddfwi}
\end{aligned} 
\end{equation}
where $\mathbf{d}_{2}^{'}=\mathbf{d}_{2} + [\mathbf{u}(\mathbf{m}_{1})- \mathbf{d}_{1}]$ can be regarded as a new dataset adjusted by the residual data of the baseline inversion. This adjustment allows unexplained data (those not fit by the earth model) in the baseline survey to be disregarded in inversions of monitoring survey data. With this definition one can use standard FWI algorithms to minimize equation (\ref{eq:ddfwi}) and obtain the optimal model $\mathbf{m}_{2}$. The time-lapse change can finally be obtained by $\delta\mathbf{m}=\mathbf{m}_{2}-\mathbf{m}_{1}$.

\subsection{Bayesian time-lapse full waveform inversion}

Bayesian inference solves inverse problems by updating a prior pdf $p(\mathbf{m})$ with new information contained in the data to construct the posterior pdf $p(\mathbf{m}|\mathbf{d})$. According to Bayes' theorem,
\begin{equation}
    p(\mathbf{m}|\mathbf{d}) = \frac{p(\mathbf{d}|\mathbf{m})p(\mathbf{m})}{p(\mathbf{d})}
\label{eq:Bayes}
\end{equation}
where $p(\mathbf{d}|\mathbf{m})$ is the \textit{likelihood} which represents the probability of observing data $\mathbf{d}$ given model $\mathbf{m}$, and $p(\mathbf{d})$ is a normalization factor called the \textit{evidence}. A Gaussian distribution is usually used to represent data uncertainties in the likelihood function, so
\begin{equation}
	p(\mathbf{d}|\mathbf{m}) \propto \mathrm{exp} \big[-\frac{1}{2}(\mathbf{d}-\mathbf{u}(\mathbf{m}))^{\mathrm{T}} \boldsymbol{\Sigma}^{-1} (\mathbf{d}-\mathbf{u}(\mathbf{m})) \big]
	\label{eq:likelihood}
\end{equation}
where $\boldsymbol{\Sigma}$ is a covariance matrix which is often assumed to be diagonal in practice.

Similarly as in standard time-lapse full waveform inversion, one can perform Bayesian inversion for each dataset $\mathbf{d}_{1}$ and $\mathbf{d}_{2}$ separately, and calculate the probability distribution $p(\delta\mathbf{m})$ using the results obtained. This can be achieved by randomly generating or selecting pairs of samples from the two posterior distribution $p(\mathbf{m}_{1}|\mathbf{d}_{1})$ and $p(\mathbf{m}_{2}|\mathbf{d}_{2})$, and computing the difference between each pair which can then be regarded as a sample of distribution $p(\delta\mathbf{m})$. The two inversions can be performed independently, or one can use the posterior samples obtained in the baseline inversion as the starting point for the monitoring inversion. Since in most Bayesian inference methods in theory the results do not depend on starting models, the two methods should produce the same results. However, by using the second strategy the burn-in period required in McMC-like methods (also sSVGD) can be significantly reduced. In this study we regard both of the above two methods as \textit{separate Bayesian inversions}, and adopt the second method in our examples below to reduce the computational cost. Since the two inversions are conducted separately, the mean and variance of $\delta\mathbf{m}$ have the following form:
\begin{subequations}
\begin{align}
	\mathrm{mean}(\delta\mathbf{m}) &= \mathrm{mean}(\mathbf{m}_{2}) - \mathrm{mean}(\mathbf{m}_{1}) \label{eq:separate_mean} \\
	\mathrm{var}(\delta\mathbf{m}) &= \mathrm{var}(\mathbf{m}_{2}) + \mathrm{var}(\mathbf{m}_{1}) \label{eq:separate_var}
\end{align}
\end{subequations}
The uncertainty of $\delta\mathbf{m}$ obtained using this method is therefore higher than that of model $\mathbf{m}_{1}$ or $\mathbf{m}_{2}$ itself. This is because the separate Bayesian inversion strategy contains an implicit assumption that the uncertainties in $\mathbf{m}_{1}$ and $\mathbf{m}_{2}$ are not correlated. This seems unlikely to be realistic: any particular earth structure at the time of the baseline survey estimated by model $\mathbf{m}_{1}$, is likely to affect uncertainties in $\mathbf{m}_{2}$ in the sense that we would expect these uncertainties to change if a different baseline earth structure was true and estimated in $\mathbf{m}_{1}$. If this correlation was taken into account we would expect the overall uncertainty on the model differences between the two surveys to decrease. Given that the magnitude of time-lapse change is usually much smaller than that of either model, we would expect that the uncertainty estimate using the above method would be less valuable in practice.

Instead of performing the two inversions separately, one can invert the two datasets simultaneously to obtain the joint distribution of model $\mathbf{m}_{1}$ and $\mathbf{m}_{2}$, that is
\begin{equation}
	p(\mathbf{m}_{1},\mathbf{m}_{2}|\mathbf{d}_{1},\mathbf{d}_{2}) = \frac{p(\mathbf{d}_{1},\mathbf{d}_{2}|\mathbf{m}_{1},\mathbf{m}_{2})
	 p(\mathbf{m}_{1})p(\mathbf{m}_{2})} {p(\mathbf{d}_{1},\mathbf{d}_{2})}
 \label{eq:joint_m1m2}
\end{equation}
This equation is still consistent with $\mathbf{m}_{1}$ and $\mathbf{m}_{2}$ being independent since in that case $p(\mathbf{d}_{1},\mathbf{d}_{2}|\mathbf{m}_{1},\mathbf{m}_{2})$ can be written as $p(\mathbf{d}_{1},\mathbf{d}_{2}|\mathbf{m}_{1},\mathbf{m}_{2})=p(\mathbf{d}_{1}|\mathbf{m}_{1})p(\mathbf{d}_{2}|\mathbf{m}_{2})$. However, given that $\mathbf{m}_{2}=\mathbf{m}_{1} + \delta\mathbf{m}$ for a change in the earth structure $\delta\mathbf{m}$, we can instead invert for the joint distribution of $\mathbf{m}_{1}$ and $\delta\mathbf{m}$:
\begin{equation}
	p(\mathbf{m}_{1},\delta\mathbf{m}|\mathbf{d}_{1},\mathbf{d}_{2}) = \frac{p(\mathbf{d}_{1},\mathbf{d}_{2}|\mathbf{m}_{1},\delta\mathbf{m})
		p(\mathbf{m}_{1})p(\mathbf{\delta m})} {p(\mathbf{d}_{1},\mathbf{d}_{2})}
	\label{eq:joint_m1dm}
\end{equation}
where $p(\delta\mathbf{m})$ is the prior distribution of $\delta\mathbf{m}$. In this way, one can impose prior information on $\delta\mathbf{m}$ by taking into account the fact that time-lapse changes are often small in practice, which therefore correlates estimates of $\mathbf{m}_{1}$ and $\mathbf{m}_{2}$ and potentially produces more accurate model change and uncertainty estimates. We refer to this method as \textit{joint Bayesian inversion}. Note that if we assume that the baseline model $\mathbf{m}_{1}$ is known, the above equation reduces to a form where we solve for the posterior distribution of $\delta \mathbf{m}$ only \citep{kotsi2020uncertainty}.
 
\subsection{Stochastic Stein variational gradient descent (sSVGD)}
To solve Bayesian inverse problems in equation (\ref{eq:Bayes}) or (\ref{eq:joint_m1dm}), we use a specific method called Stochastic Stein variational gradient descent (sSVGD) which combines Monte Carlo and variational inference methods \citep{gallego2018stochastic}. The method simulates a Markov process using a stochastic differential equation (SDE):
\begin{equation}
	\mathrm{d}\mathbf{z} = \mathbf{f}(\mathbf{z})\mathrm{d}t + \sqrt{2\mathbf{D}(\mathbf{z})}d\mathbf{W}(t)
	\label{eq:sde}
\end{equation}
where $\mathbf{z} \in \mathbb{R}^{m}$, $\mathbf{f}(\mathbf{z})$ is called the \textit{drift}, $\mathbf{W}(t)$ is a Wiener process and $\mathbf{D}(\mathbf{z})$ is a positive semidefinite diffusion matrix. If we denote the posterior distribution of interest from either equation (\ref{eq:Bayes}) or (\ref{eq:joint_m1dm}) as $p(\mathbf{z})$, \cite{ma2015complete} proposed a specific form of equation (\ref{eq:sde}) which gives an SDE that converges to distribution $p(\mathbf{z})$:
\begin{equation}
	\mathbf{f}(\mathbf{z}) = [\mathbf{D}(\mathbf{z})+\mathbf{Q}(\mathbf{z})] \nabla \mathrm{log}p(\mathbf{z}) +  \Gamma(\mathbf{z})
	\label{eq:sde_drift}
\end{equation}
where $\mathbf{Q}(\mathbf{z})$ is a skew-symmetric curl matrix, $\Gamma(\mathbf{z}) = \sum_{j=1}^{m} \frac{\partial}{\partial\mathbf{z}_{j}}(\mathbf{D}_{ij}(\mathbf{z}) + \mathbf{Q}_{ij}(\mathbf{z}))$, and $\nabla \mathrm{log}p(\mathbf{z})$ represents derivatives of $\mathrm{log}p(\mathbf{z})$ with respect to all variables in $\mathbf{z}$. By choosing different matrices $\mathbf{D}$ and $\mathbf{Q}$, different methods can be obtained \citep{ma2015complete}. For example, if we choose $\mathbf{D}=\mathbf{I}$ and $\mathbf{Q}=\mathbf{0}$ we obtained the stochastic gradient Langevin dynamics algorithm \citep{welling2011bayesian}. If we construct an augmented space  $\overline{\mathbf{z}}=(\mathbf{z},\mathbf{x})$ by concatenating $\mathbf{z}$ and a moment term $\mathbf{x}$, and set  $\mathbf{D}=\mathbf{0}$ and 
$\mathbf{Q}=\left[ 
{\small 
	\begin{array}{cc}
	\mathbf{0} & \mathbf{-I} \\ \mathbf{I} & \mathbf{0}
	\end{array} 
} 
\right]$, we obtain the stochastic Hamiltonian Monte Carlo method \citep{chen2014stochastic}.

The above process can be simulated numerically by discretizing equation (\ref{eq:sde}) with equation (\ref{eq:sde_drift}) over time variable $t$ using the Euler–Maruyama discretization:
\begin{equation}
	\mathbf{z}_{t+1} = \mathbf{z}_{t} + \epsilon_{t} \left[ \left(\mathbf{D}\left(\mathbf{z}_{t}\right) + \mathbf{Q}(\mathbf{z}_{t})\right)\nabla \mathrm{log}p(\mathbf{z}_{t}) + \Gamma(\mathbf{z}_{t}) \right] + \mathcal{N}(\mathbf{0},2\epsilon_{t}\mathbf{D}(\mathbf{z}_{t}))
	\label{eq:discretized_sde}
\end{equation}
where $\epsilon_{t}$ is a small step, and $\mathcal{N}(\mathbf{0},2\epsilon_{t}\mathbf{D}(\mathbf{z}_{t}))$ is a Gaussian distribution with mean $\mathbf{0}$ and covariance matrix $2\epsilon_{t}\mathbf{D}(\mathbf{z}_{t})$. Since $p(\mathbf{z}_{t})$ represents the posterior distribution in equation (\ref{eq:Bayes}), it depends implicitly on observed data $\mathbf{d}$. The gradient $\nabla \mathrm{log}p(\mathbf{z}_{t})$ can be calculated using either the full dataset or Uniformly randomly selected minibatch datasets in each step in $t$, and in either case the process converges to the posterior distribution $p(\mathbf{z})$ when $\epsilon_{t} \to 0$ and $t \to \infty$. 

sSVGD uses a set of models called particles since sSVGD moves them through parameter space. Define the set of particles as $\{\mathbf{m}_{i}: i=1,...,n\}$ where $\mathbf{m}_{i} \in \mathbb{R}^{r}$, and construct an augmented space $\mathbf{z}=(\mathbf{m}_{1}, \mathbf{m}_{2}, ..., \mathbf{m}_{n}) \in \mathbb{R}^{nr}$ by concatenating the $n$ particles. Using equation (\ref{eq:discretized_sde}) we construct a sampler that runs $n$ multiple interacting chains:   
\begin{equation}
	\mathbf{z}_{t+1} = \mathbf{z}_{t} + \epsilon_{t} [(\mathbf{D}(\mathbf{z}_{t}) + \mathbf{Q}(\mathbf{z}_{t}))\nabla \mathrm{log}p(\mathbf{z}_{t}) + \Gamma(\mathbf{z}_{t})] + \mathcal{N}(\mathbf{0},2\epsilon_{t}\mathbf{D}(\mathbf{z}_{t}))
	\label{eq:general_sgmc}
\end{equation}
where $\mathbf{D}, \mathbf{Q} \in \mathbb{R}^{nr\times nr}$ and $\nabla \mathrm{log}p, \Gamma \in \mathbb{R}^{nr}$. Define a matrix $\mathbf{K}$:
\begin{equation}
	\mathbf{K} = \frac{1}{n} \begin{bmatrix}
		k(\mathbf{m}_{1},\mathbf{m}_{1})\mathbf{I}_{r\times r} & \dots & k(\mathbf{m}_{1},\mathbf{m}_{n})\mathbf{I}_{r\times r} \\
		\vdots & \ddots & \vdots \\
		k(\mathbf{m}_{n},\mathbf{m}_{1})\mathbf{I}_{r\times r} & \dots & k(\mathbf{m}_{n},\mathbf{m}_{n})\mathbf{I}_{r\times r} 
	\end{bmatrix}
	\label{eq:matrixK}
\end{equation}
where $k(\mathbf{m}_{i},\mathbf{m}_{j})$ is a kernel function and $\mathbf{I}_{r \times r}$ is an identity matrix. Note that $\mathbf{K}$ is positive definite according to the definition of kernel functions \citep{gallego2018stochastic}. By setting $\mathbf{D}=\mathbf{K}$ and $\mathbf{Q}=\mathbf{0}$, equation (\ref{eq:general_sgmc}) becomes:
\begin{equation}
	\mathbf{z}_{t+1} = \mathbf{z}_{t} + \epsilon_{t} [\mathbf{K} \nabla \mathrm{log}p(\mathbf{z}_{t}) + \nabla \cdot \mathbf{K}]
	+ \mathcal{N}(\mathbf{0},2\epsilon_{t}\mathbf{K})
	\label{eq:stochastic_svgd}
\end{equation} 
This defines a Markov process that converges to the posterior distribution $p(\mathbf{z})= \prod_{i=1}^{n} p(\mathbf{m}_{i}|\mathbf{d})$ asymptotically for any number of particles $n$. Note that if we eliminate the noise term $\mathcal{N}(\mathbf{0},2\epsilon_{t}\mathbf{K})$ in equation (\ref{eq:stochastic_svgd}), the method becomes Stein variational gradient descent (SVGD). The sSVGD algorithm is therefore a stochastic gradient McMC method that uses SVGD gradients \citep{gallego2018stochastic}.

Equation (\ref{eq:stochastic_svgd}) requires that we generate samples from the distribution $\mathcal{N}(\mathbf{0},2\epsilon_{t}\mathbf{K})$, which can be computationally expensive because the matrix $\mathbf{K}$ is potentially large. To perform this more efficiently we define a block diagonal matrix $\mathbf{D}_{\mathbf{K}}$
\begin{equation}
	\mathbf{D}_{\mathbf{K}} = \frac{1}{n} 
	\left[ \renewcommand\arraystretch{0.3}
	\begin{array}{ccc}
		\overline{\mathbf{K}} & & \\
		& \ddots & \\
		& & \overline{\mathbf{K}}
	\end{array}
	\right]
	\label{eq:matrixDK}
\end{equation}
where $\overline{\mathbf{K}}$ is a $n \times n$ matrix with $\overline{\mathbf{K}}_{ij} = k(\mathbf{m}_{i},\mathbf{m}_{j})$. Note that the matrix $\mathbf{D}_{\mathbf{K}}$ can be constructed from $\mathbf{K}$ using $\mathbf{D}_{\mathbf{K}} = \mathbf{P}\mathbf{K}\mathbf{P}^{\mathrm{T}}$ where $\mathbf{P}$ is a permutation matrix
\begin{equation}
	\mathbf{P} = 
	\left[\arraycolsep=2pt \def\arraystretch{0.2}
	\begin{array}{clc|clc|c|clc}
		1 & & & & & & & & & \\
		& & & 1 & & & & & &  \\
		& & & & & & \ddots & & & \\
		& & & & & & & 1 & & \\
		\hline
		& 1 & & & & & & & & \\
		& & & & 1 & & & & &  \\
		& & & & & & \ddots & & & \\
		& & & & & & & & 1 & \\
		\hline
		& \ddots & & & \ddots & & \ddots & & \ddots & \\
		\hline
		& & 1 & & & & & & & \\
		& & & & & 1 & & & &  \\
		& & & & & & \ddots & & & \\
		& & & & & & & & & 1 \\
	\end{array}
	\right]
\end{equation}
The action of $\mathbf{P}$ on a vector $\mathbf{z}$ rearranges the order of the vector elements from the basis where particles are concatenated sequentially to the basis where the first coordinates of all the particle are listed, then the second, etc. With this definition, a sample $\boldsymbol{\eta}$ can be generated more efficiently from $\mathcal{N}(\mathbf{0},2\epsilon_{t}\mathbf{K})$ using
\begin{equation}
	\begin{aligned}
		\boldsymbol{\eta} &\sim \mathcal{N}(\mathbf{0},2\epsilon_{t}\mathbf{K}) \\
		&\sim \sqrt{2\epsilon_{t}} \mathbf{P}^{\mathrm{T}}\mathbf{P}\mathcal{N}(\mathbf{0},\mathbf{K}) \\
		&\sim \sqrt{2\epsilon_{t}} \mathbf{P}^{\mathrm{T}} \mathcal{N}(\mathbf{0},\mathbf{D}_{\mathbf{K}}) \\
		&\sim \sqrt{2\epsilon_{t}} \mathbf{P}^{\mathrm{T}} \mathbf{L}_{\mathbf{D}_\mathbf{K}}  \mathcal{N}(\mathbf{0},\mathbf{I})
	\end{aligned}
	\label{eq:noise_term}
\end{equation}
where $\mathbf{L}_{\mathbf{D}_\mathbf{K}}$ is the lower triangular Cholesky decomposition of matrix $\mathbf{D}_{\mathbf{K}}$, which can be calculated easily as only the lower triangular Cholesky decomposition of matrix $\overline{\mathbf{K}}$ is required by equation (\ref{eq:matrixDK}). In practice the number of particles $n$ is usually sufficiently modest that the decomposition of $\overline{\mathbf{K}}$ is computationally negligible. We can thus use equation (\ref{eq:stochastic_svgd}) to generate samples of the posterior distribution. 

\begin{figure}
	\includegraphics[width=1.\linewidth]{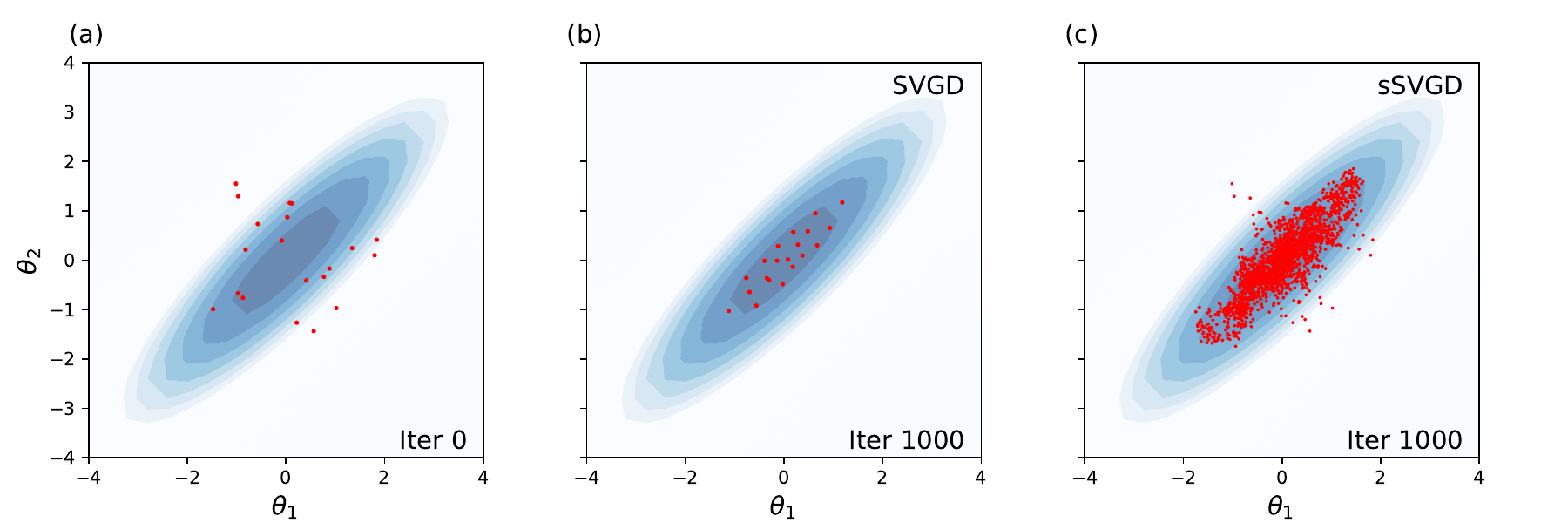}
	\caption{An example which uses SVGD and sSVGD to sample a bivariate Gaussian distribution (blue shades). Red dots show (\textbf{a}) the initial 20 particles, and the samples obtained using (\textbf{b}) SVGD and (\textbf{c}) sSVGD after 1,000 iterations.}
	\label{fig:bivariate_example}
\end{figure}

Figure \ref{fig:bivariate_example} shows an example in which the sSVGD algorithm is used to generate samples from a bivariate Gaussian distribution. It compares the results to those of SVGD after the same number of iterations. The main practical difference between the algorithms is that sSVGD generates many more samples of the distribution than SVGD, since the particle values from every iteration (potentially after some burn-in period) constitute valid samples. In geophysics sSVGD has already been used to solve 3D FWI problems \citep{zhang20233}; in this study we test the method in the context of solving time-lapse imaging problems by sampling the distributions in equations (\ref{eq:Bayes}) and (\ref{eq:joint_m1dm}).

\section{Results}
\subsection{Experimental setup}

\begin{figure}
	\includegraphics[width=1.\linewidth]{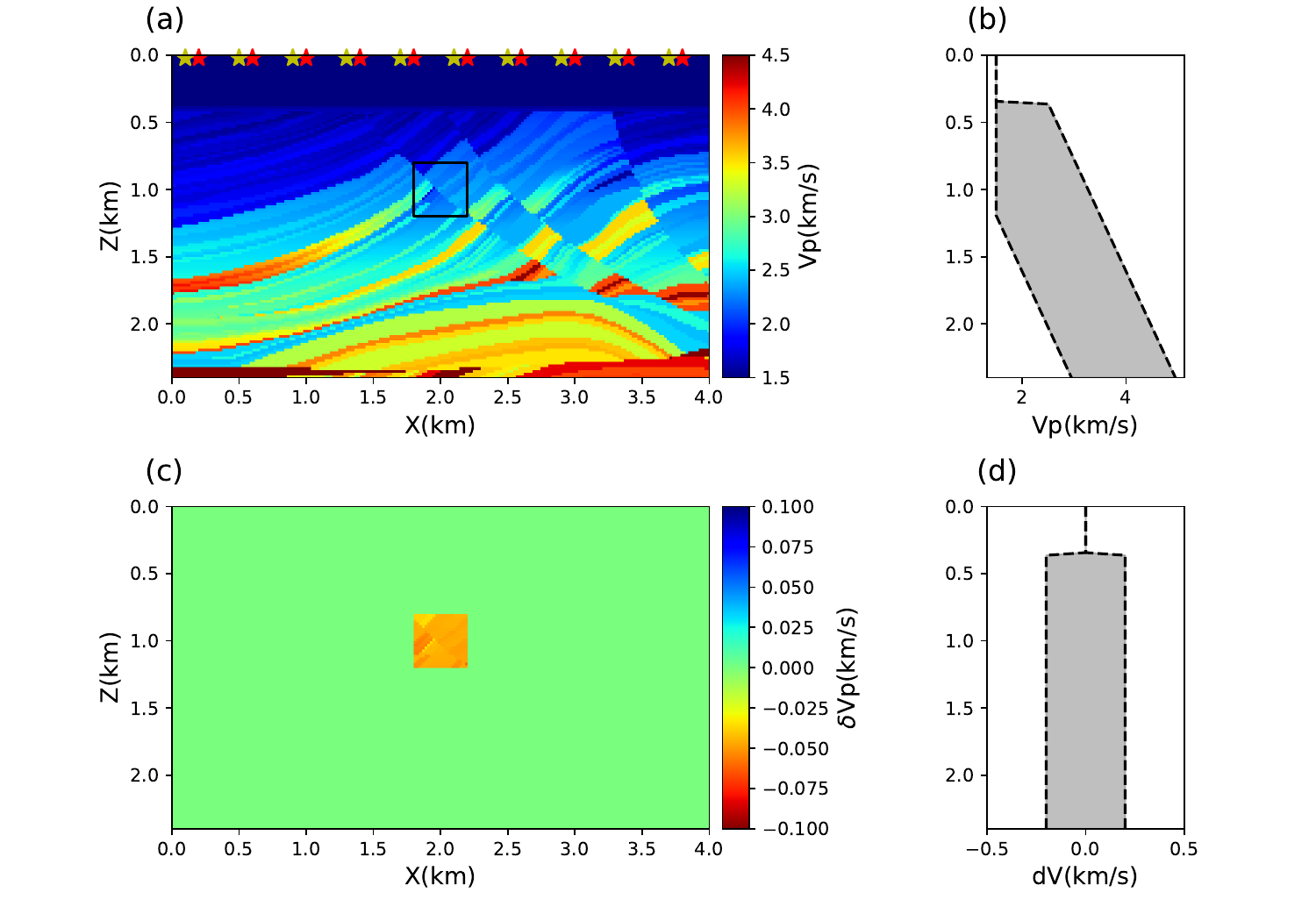}
	\caption{(\textbf{a}) The true velocity model at the time of the baseline survey, and the acquisition geometry used in this study. Red stars denote the source locations in the baseline survey while yellow stars show the perturbed locations in the monitoring survey. 200 receivers are equally spaced at the seabed at 360 m depth (not shown). (\textbf{b}) The prior distribution of velocity: a Uniform distribution with an interval of 2 km/s, other than above 1.2 km depth where an extra lower bound of 1.5 km/s is also imposed to ensure rock velocity is higher than the velocity in water. (\textbf{c}) The true time-lapse velocity change in the monitoring survey. (\textbf{d}) The prior distribution of velocity change which is set to be a Uniform distribution over an interval of $\pm$0.2 km/s.}
	\label{fig:model_perturbation_prior}
\end{figure}

To understand the robustness and behavior of each method, we set up a synthetic time-lapse experiment using a part of the Marmousi model to represent the true baseline model \cite[Figure \ref{fig:model_perturbation_prior}a,][]{martin2006marmousi2}. To represent the true time-lapse model we reduce the velocity of a small square area in the baseline model by 2\% (Figure \ref{fig:model_perturbation_prior}a and \ref{fig:model_perturbation_prior}c). We choose a square area since this discriminates geometrically correct imaging results from errors, since some of the latter are shown below to follow geological strata and hence to look realistic. Both baseline and time-lapse models are parameterized using a regular 200 $\times$ 100 grid with a spacing of 20 m. Ten sources are located at 20 m water depth (red stars in Figure \ref{fig:model_perturbation_prior}), and 200 equally spaced receivers are located on the seabed at 360m water depth across the horizontal extent of the model. Since it is not possible to repeat exactly the same acquisition geometries in time-lapse seismic surveys, and attempts to do so usually incur significant cost \citep{beasley1999repeatability, yang2015double, calvert20054d}, we study performance of the different methods when the source locations are repeated and when they are perturbed by 100 m in the monitoring survey (yellow stars in Figure \ref{fig:model_perturbation_prior}). In both cases we assume that the locations of the source positions used in each survey are known. All waveform data are simulated using a time-domain finite difference method with a Ricker wavelet of 10 Hz central frequency, and we added 1\% uncorrelated Gaussian noise to the data. For all inversions the gradients of the misfit (likelihood) function with respect to wave velocity in each cell are calculated using the adjoint method \citep{tarantola1988theoretical, tromp2005seismic, fichtner2006adjoint, plessix2006review}.

For the prior information on absolute seismic velocity we use a Uniform distribution over an interval of 2 km/s at each depth (Figure \ref{fig:model_perturbation_prior}b). To ensure that the rock velocity is higher than the velocity in water we impose an additional lower bound of 1.5 km/s. Given that time-lapse changes in seismic velocity are usually much smaller than the velocity itself, we use a Uniform distribution between -0.2 km/s and 0.2 km/s (Figure \ref{fig:model_perturbation_prior}d) for the prior information $p(\delta\mathbf{m})$ in equation (\ref{eq:joint_m1dm}). Note that no prior information is imposed directly on the model difference $\delta\mathbf{m}$ in the separate Bayesian inversion strategy.

\subsection{Exactly repeated acquisition geometry}
\begin{figure}
	\includegraphics[width=1.\linewidth]{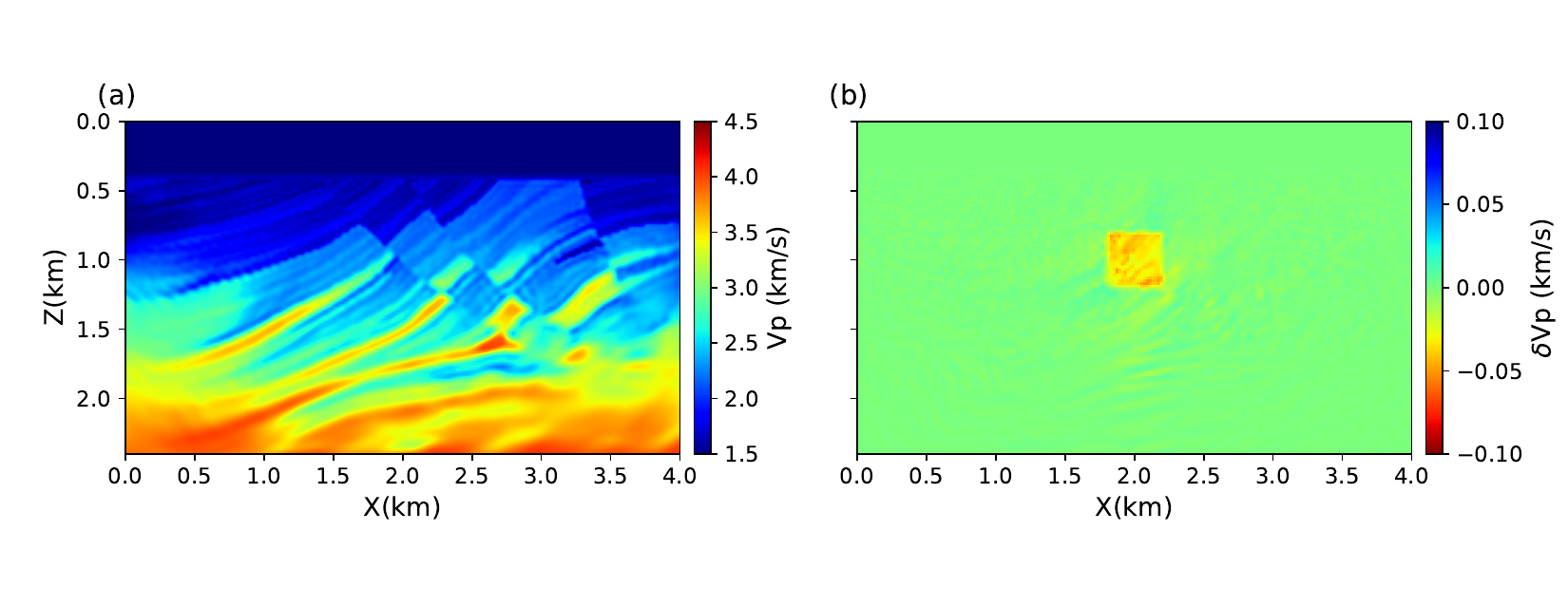}
	\caption{(\textbf{a}) Baseline velocity model obtained using standard linearised FWI.  (\textbf{b}) Time-lapse velocity change obtained using the double difference method with identical source locations in the baseline and monitoring survey.}
	\label{fig:dd_results}
\end{figure}

We first perform time-lapse studies with the acquisition geometry repeated identically in the baseline and monitor surveys. The standard double difference FWI method requires a good baseline model to obtain accurate velocity changes \citep{asnaashari2015time}. To attempt to achieve this we adopt a multiscale FWI strategy \citep{bunks1995multiscale} in the baseline inversion: we first invert for a long wavelength model using low frequency data simulated using a Ricker wavelet of 4 Hz central frequency. The initial model in this low frequency inversion is set to be laterally-constant with velocity equal to the average velocity of the prior distribution in Figure \ref{fig:model_perturbation_prior}b, and the range of models in Figure 2b is also imposed as a set of constraints on velocities at each depth. The resulting long wavelength model serves as the starting model for the inversion using higher frequency data (10 Hz wavelet). For both inversions we use the LBFGS method \citep{liu1989limited} to minimize misfit functions as in equation (\ref{eq:fwi_loss}), in which the control parameter of the regularization term is selected by trial and error. Figure \ref{fig:dd_results}a shows the obtained baseline model which provides an accurate estimate of the true model. Note that because of the low data sensitivity around the bottom and edges of the model caused by acquisition geometry limitations, velocity structures in these areas exhibit larger errors. 

We then use this baseline model to conduct double difference FWI by minimizing the misfit function in equation (\ref{eq:ddfwi}). The obtained time-lapse changes are shown in Figure \ref{fig:dd_results}b. The results demonstrate that the double difference method can obtain reasonably accurate estimates of the true velocity changes, as found in previous studies \citep{watanabe2004differential, denli2009double, zheng2011strategies, asnaashari2015time, yang2015double}. However, the method cannot provide accurate uncertainty estimates for those velocity changes since it only accounts for linearised physics relating model parameters and data \citep{smith2013uncertainty, zhang20183}.

To quantify uncertainties in velocity changes we perform time-lapse studies using the above two Bayesian methods. For the separate Bayesian inversion we adopt the strategy which uses the particles of the baseline inversion as the starting point for the monitoring survey inversion as this 'warm start' procedure has been demonstrated to be more effective than two independent inversions in linearised methods \citep{zheng2011strategies, asnaashari2015time}, and was shown to be similarly effective when data of increasingly high frequency were added to an FWI solution found by SVGD \citep{zhang2021bayesiana}. The prior distributions are set to be the same for both baseline and monitoring inversions, equal to the Uniform distribution in Figure \ref{fig:model_perturbation_prior}b. In the baseline inversion we randomly generate 20 samples from the prior distribution as the initial particles, which are then updated using equation (\ref{eq:stochastic_svgd}) for 4,000 iterations after an additional burn-in period of 2,000. For the kernel function in equation (\ref{eq:matrixK}) we employ a commonly-used radial basis function,
\begin{equation}
	k(\mathbf{m}_{1},\mathbf{m}_{2}) = \mathrm{exp} [- \frac{\Vert \mathbf{m}_{1}-\mathbf{m}_{2} \Vert^{2}}{2h^{2}}]
	\label{eq:rbf}
\end{equation}
where $h$ is a scale factor that controls the intensity of interaction between two particles based on their distance apart. As suggested by previous studies \citep{liu2016stein, zhang2020seismic}, we choose $h$ to be $\tilde{d} / \sqrt{2\mathrm{log}n}$ where $\tilde{d}$ is the median of pairwise distances between all particles and $n$ is the number of particles. To reduce the memory and storage cost we only retain every tenth sample after a burn-in period of 2,000 iterations, which results in a total of 8,000 samples. Those samples are then used to calculate statistics (mean and standard deviation) of the posterior distribution.

\begin{figure}
	\includegraphics[width=1.\linewidth]{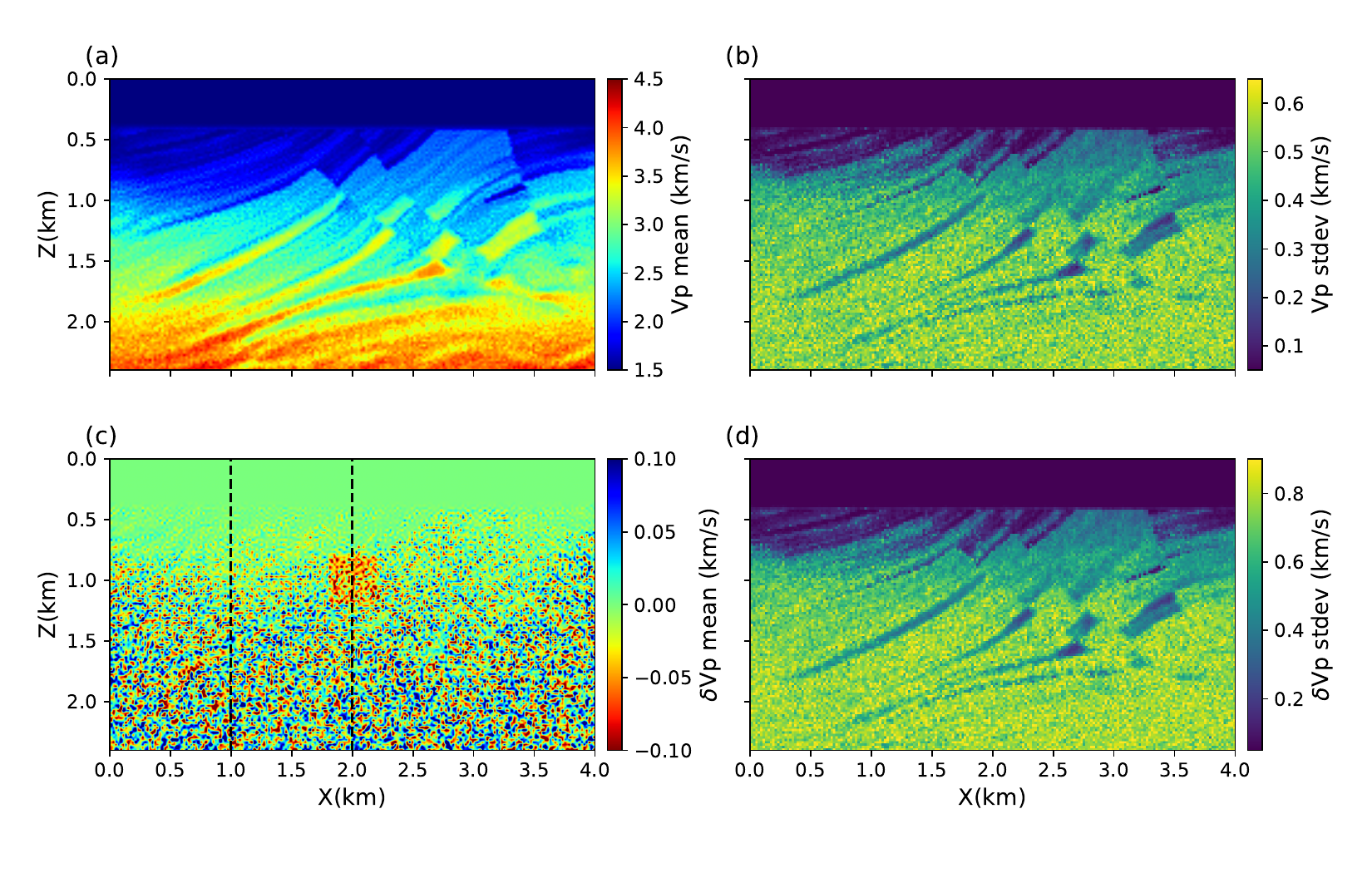}
	\caption{(\textbf{a}) The mean and (\textbf{b}) standard deviation of velocity obtained using sSVGD in the baseline survey. (\textbf{c}) The mean and (\textbf{d}) standard deviation of velocity change obtained using separate Bayesian inversions with identical source locations in the baseline and monitoring survey. The dashed black lines show well log locations referred to in the main text. Abbreviation stdev stands for standard deviation.}
	\label{fig:separate_results}
\end{figure}

Figure \ref{fig:separate_results}a and b show the mean and standard deviation models obtained in the baseline inversion. Although the inversion is performed using the high frequency data directly, the mean model still provides an accurate estimate of the true model, similarly to the linearised inversion which uses the extra low frequency data set described above. Again similarly to the linearised inversion, the structure in the bottom and edges differ from the true model because of low sensitivity. Note that the mean model shows pixel-scale randomness which reflects the true uncertainty of neighbouring pixels since there is nothing in the problem setup that prefers smooth models, and neither does seismic waveform data. The model obtained using the linearised method is much smoother because smoothness was imposed as additional regularization. Overall the standard deviation map shows similar geometries to the mean, which has also been found in previous studies \citep{zhang2021bayesiana, gebraad2020bayesian, zhang2020variational}. In addition, the results show higher uncertainties at large depths ($>$ 1.2km) because of reduced data sensitivity, which is also consistent with results obtained using the SVGD method \citep{zhang2021bayesiana}.

For the monitoring inversion we restart the sampling in the above sSVGD from the final 20 particles using the new dataset $\mathbf{d}_{2}$, and continue for another 2,000 iterations. No burn-in period is specified for this inversion as the starting models are supposed to be close to the true model. In addition we only retain every fifth sample of particle values so that the total number of samples used is the same as that in the baseline inversion. To obtain samples of the time-lapse change, we randomly select pairs of samples from the two sets of model samples obtained in the baseline and monitoring inversion, and calculate the time-lapse change using $\delta\mathbf{m}=\mathbf{m}_{2}-\mathbf{m}_{1}$. The statistics of the posterior distribution of time-lapse change can then be computed.

Figure \ref{fig:separate_results}c and d show the mean and standard deviation maps of time-lapse changes. As in the double difference inversion, the mean map clearly shows the outline of true velocity change.  However, there are additional small scale structures (a few pixels in size) in the results, which may reflect the true uncertainty in the problem itself, or may exist because the algorithm has not fully converged given that this is a high dimensional problem ($r=$24,000). Either way, those structures do not affect the overall interpretation of the results from a geological point of view. The standard deviation map shows almost the same structure as that obtained in the baseline inversion except that the magnitude is much higher. This is because the two inversions are conducted separately, and the variance of the time-lapse change is the summation of the variances of velocity obtained in each inversion (equation \ref{eq:separate_var}). As a result, the standard deviation model is not particularly useful in practice as the magnitude of uncertainty is far higher than that of the time-lapse change itself.   

\begin{figure}
\includegraphics[width=1.\linewidth]{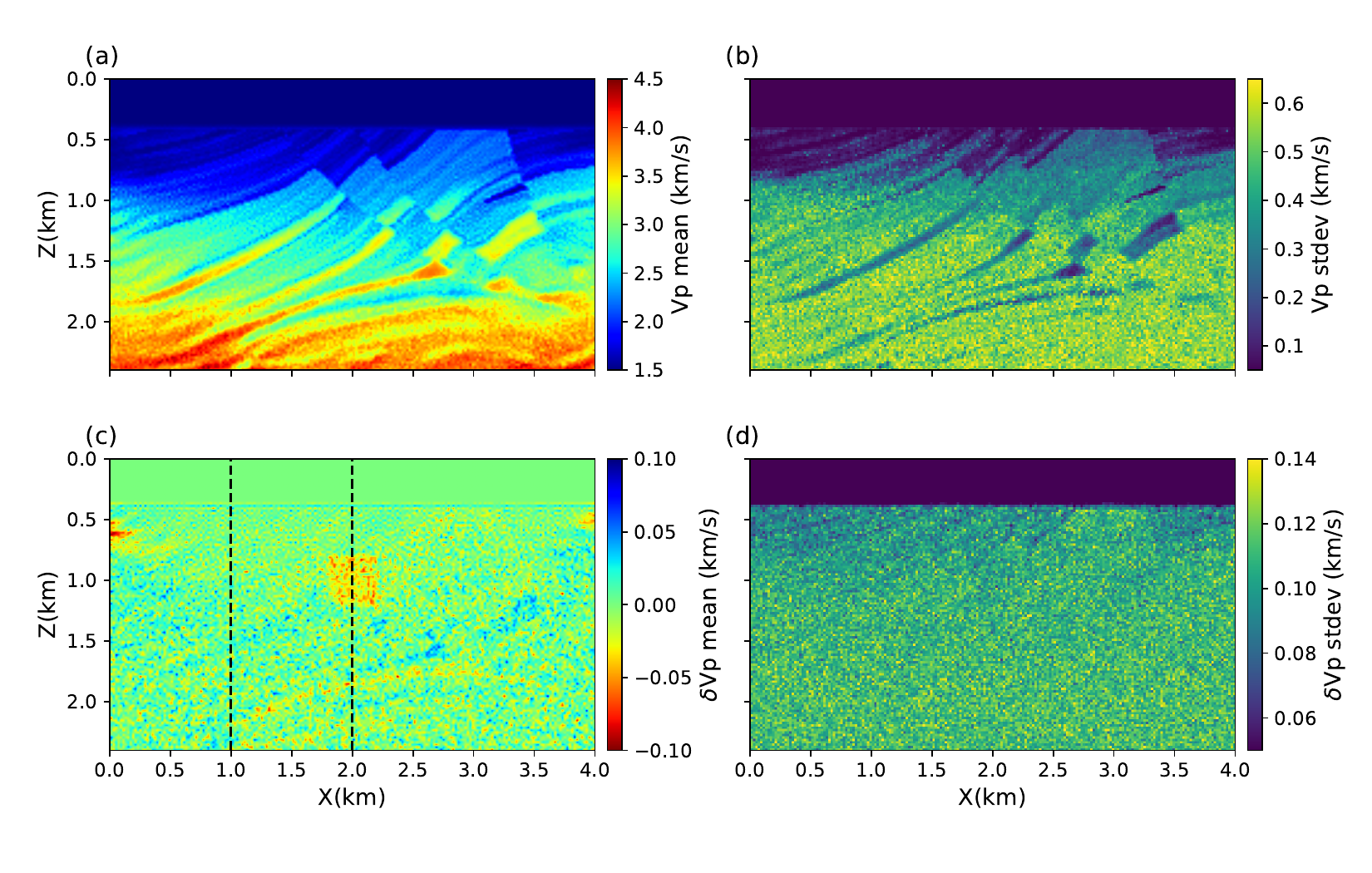}
\caption{The mean and standard deviation of velocity (\textbf{top}) and velocity change (\textbf{bottom}) obtained using the joint Bayesian inversion strategy. Key as in Figure \ref{fig:separate_results}.}
\label{fig:joint_results}
\end{figure}

In the joint Bayesian inversion we use the same prior distribution for velocity (Figure \ref{fig:model_perturbation_prior}b) and add additional prior information about the velocity change (Figure \ref{fig:model_perturbation_prior}c). Similarly to above, we generate 20 particles from the prior distribution and update them using equation (\ref{eq:stochastic_svgd}) for 4,000 iterations with an additional burn-in period of 4,000. Only every tenth sample is retained which results in a total of 8,000 samples. Other settings of the sSVGD method are kept the same as in the separate inversion strategy above. Finally, statistics of the posterior distribution $p(\mathbf{m}_{1},\delta\mathbf{m}|\mathbf{d}_{1},\mathbf{d}_{2})$ are computed using the samples obtained.

Figure \ref{fig:joint_results} shows the results obtained in the joint inversion. The mean and standard deviation of velocity obtained for the baseline period (Figure \ref{fig:joint_results}a and b) show almost the same structures as those obtained in the separate Bayesian inversions (Figure \ref{fig:separate_results}a and b). For example, the mean also represents a good estimate of the true model and the standard deviation shows similar geometrical features to those on the mean map. Thus, although the number of parameters is doubled in the joint inversion compared to the separate inversion strategy, the method still provides good estimates of the baseline model.

Similarly to the results obtained above, the mean model of velocity change provides a reasonable estimate of the true velocity change (Figure \ref{fig:joint_results}c). There are also small scale structures in the mean model as in the separate Bayesian inversion, which probably have similar origins. However, the magnitudes of those structures are much smaller than those from the separate Bayesian inversions. Note that there are also some negative velocity changes at the edges around the depth of 0.6 km and some geological structures close to the bottom around depth of 2.0 km that are associated with similar structures in the velocity model. This is probably because the data sets cause velocity and velocity change to be correlated with each other, and consequently uncertainty in velocity can introduce uncertainties to the velocity change. The standard deviation model indicates that the uncertainty estimates from joint inversion are almost an order of magnitude smaller than those obtained using separate inversions because of the additional prior information imposed on the velocity change. Similarly to the standard deviation of velocity, the uncertainty of velocity change is smaller in shallow parts ($<$ 1.0km) and larger in deeper parts ($>$ 1.0 km) of the model. Note that because of coupling between velocity and velocity change, the magnitude of uncertainty in the deeper part is actually similar to that of the prior distribution (0.12 km/s). This indicates that the uncertainty in the baseline model can have a high impact on the uncertainty in velocity changes. Nevertheless, compared to the results obtained using separate inversions, the joint inversion strategy produces more accurate velocity changes and more reliable uncertainty estimates.

\subsection{Perturbed acquisition geometry}
\begin{figure}
	\includegraphics[width=1.\linewidth]{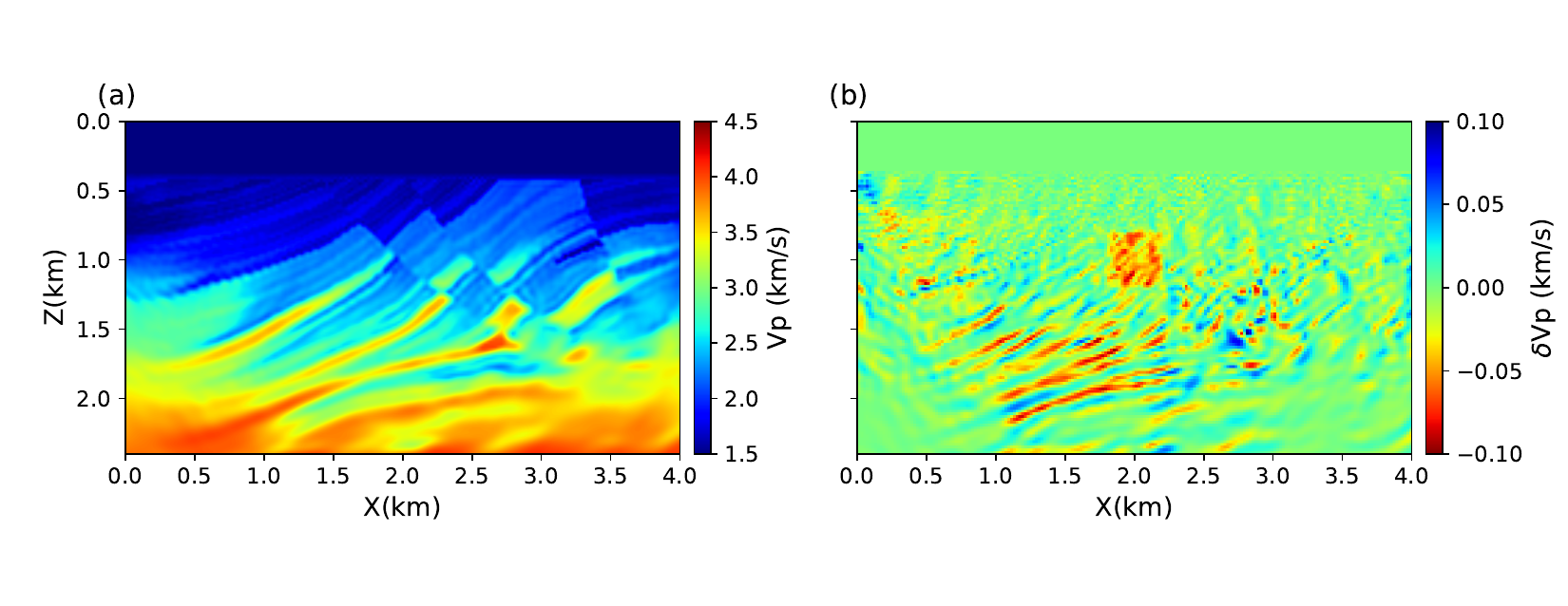}
	\caption{(\textbf{a}) The baseline velocity model obtained using the standard linearised method. (\textbf{b}) The time-lapse velocity change obtained using the double difference method with perturbed source locations in the monitoring survey (yellow stars in Figure \ref{fig:model_perturbation_prior}a).}
	\label{fig:dd_perturbedsource_results}
\end{figure}

We now study the performance of the suite of methods in the case in which source positions are perturbed in the monitoring survey (yellow stars in Figure \ref{fig:model_perturbation_prior}a). Although it is possible to use time-lapse binning or data interpolation to emulate repeatable data acquisition \citep{asnaashari2015time}, in this study none of these procedures is performed because of the sparseness of the source positions which make these procedures inaccurate, and also because our purpose is to study the reliability of different inversion methods under different geometries. In the double difference inversion we use the same baseline model as above and follow the same procedure to minimize the misfit function in equation (\ref{eq:ddfwi}), accounting for the different source locations in baseline and monitoring surveys (i.e., assuming that these are known). The obtained time-lapse change is shown in Figure \ref{fig:dd_perturbedsource_results}. Although the shape of the true velocity change can be observed in the results, there are many additional structures which have similar magnitudes to the true velocity change but do not represent any real changes. Since these structures follow geological strata, they can certainly bias dynamic interpretations of the observed changes. We therefore conclude that double difference FWI generates significantly biased results in the case of perturbed acquisition geometries, even if the perturbed source locations are known, a result that has also been found in previous studies \citep{asnaashari2015time, yang2015double}.
\begin{figure}
	\includegraphics[width=1.\linewidth]{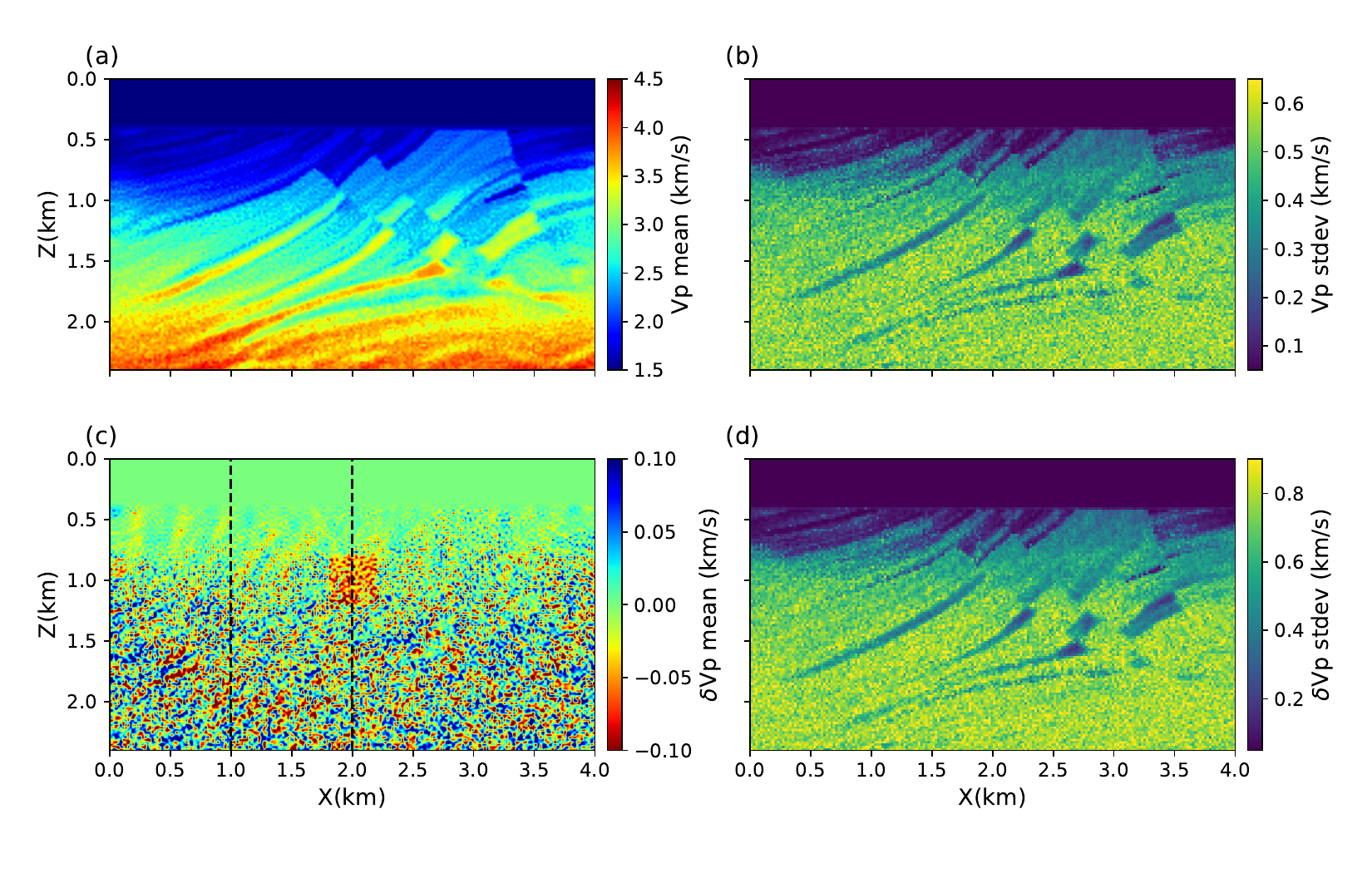}
	\caption{The mean and standard deviation of velocity (\textbf{top}) and velocity change (\textbf{bottom}) obtained using the separate Bayesian inversion strategy with perturbed source locations. Key as in Figure \ref{fig:separate_results}.}
	\label{fig:separate_perturbedsource_results}
\end{figure}
\begin{figure}
	\includegraphics[width=1.\linewidth]{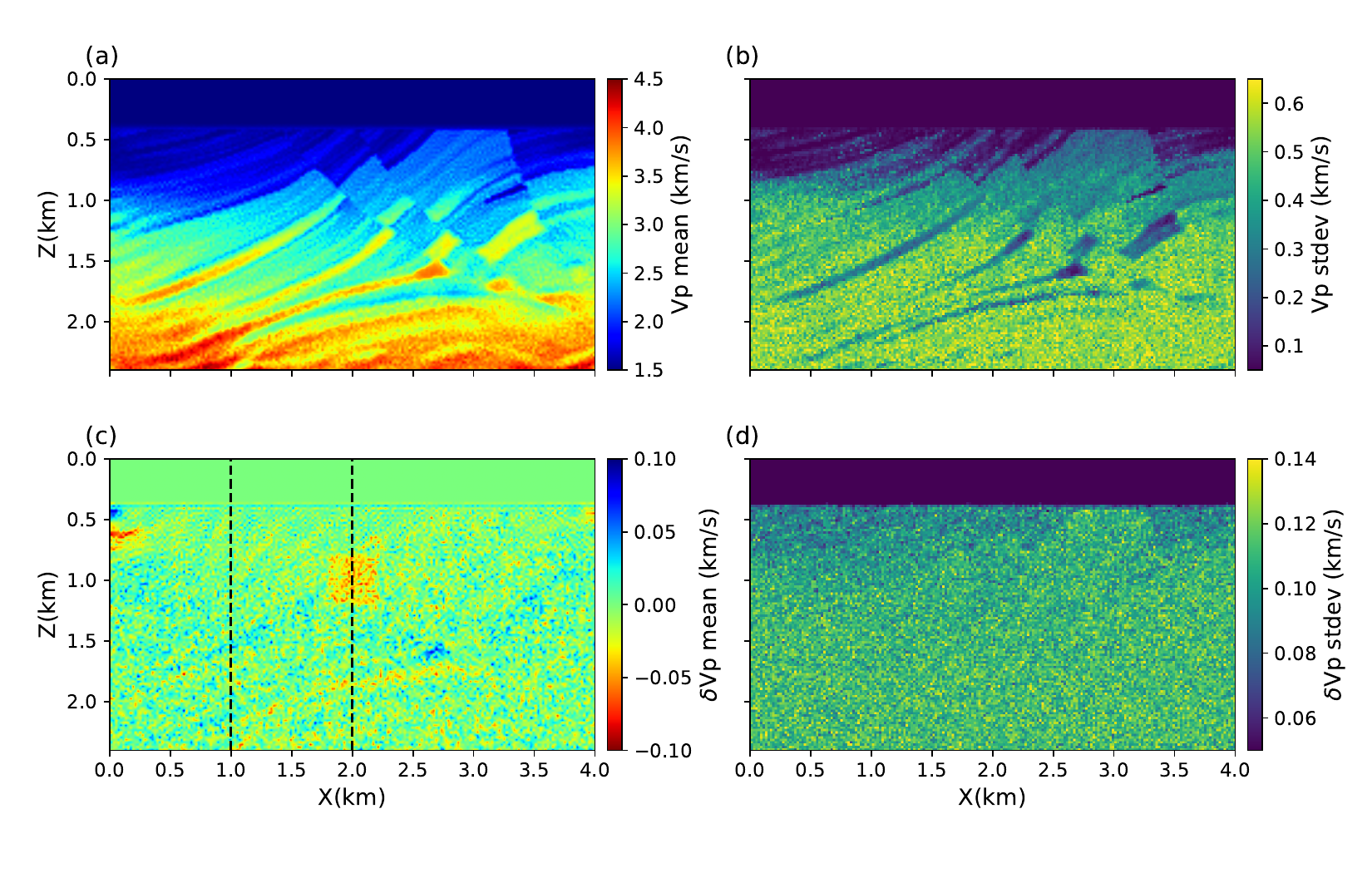}
	\caption{The mean and standard deviation of velocity (\textbf{top}) and velocity change (\textbf{bottom}) obtained using the joint Bayesian inversion strategy with perturbed source locations. Key as in Figure \ref{fig:separate_results}.}
	\label{fig:joint_perturbedsource_results}
\end{figure}

\begin{figure}
	\includegraphics[width=1.\linewidth]{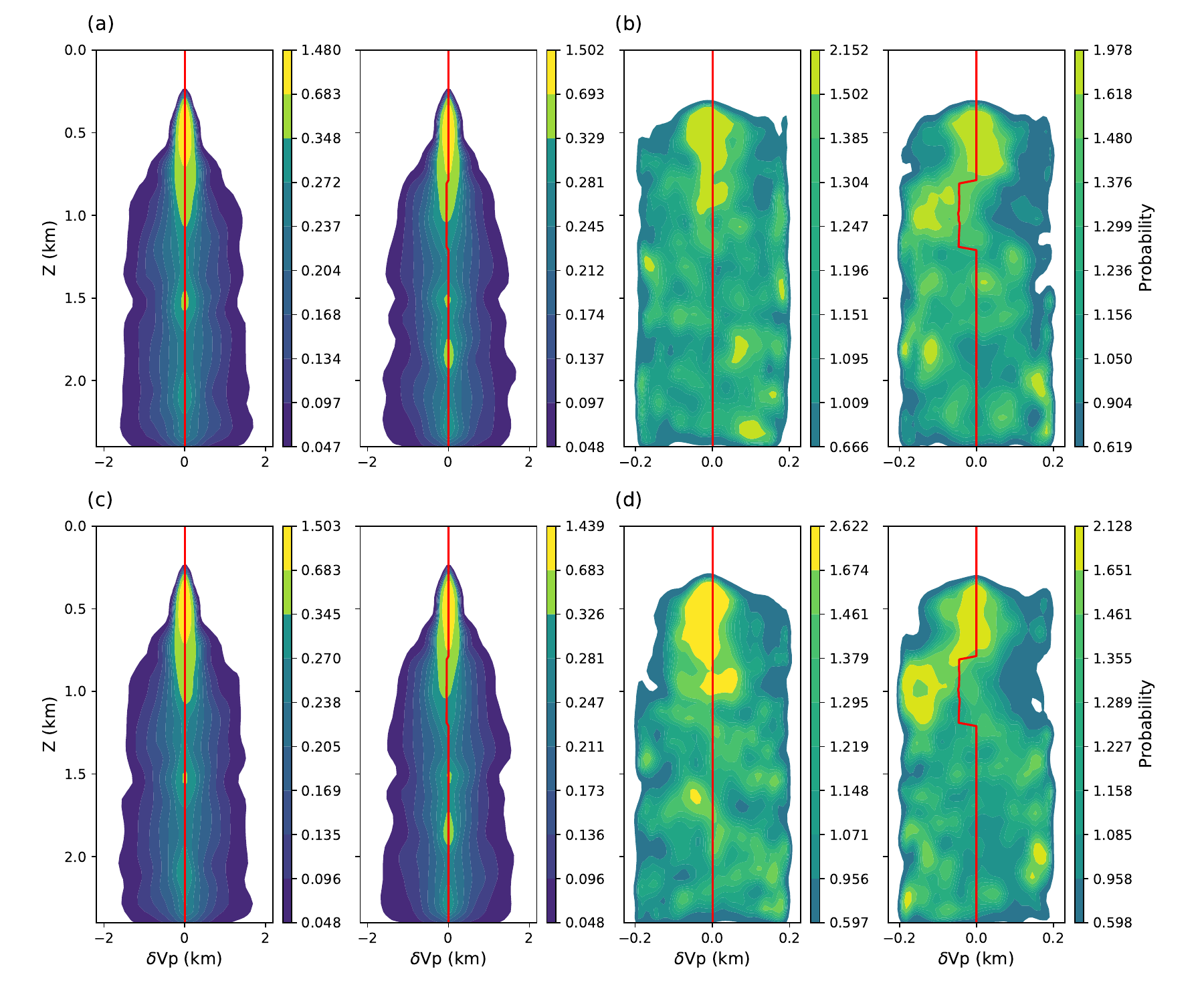}
	\caption{The marginal distributions of velocity change at two well locations (black dashed line in Figure \ref{fig:separate_results}, \ref{fig:joint_results}, \ref{fig:separate_perturbedsource_results} and \ref{fig:joint_perturbedsource_results}) obtained using (\textbf{a}, \textbf{c}) separate Bayesian inversion and (\textbf{b}, \textbf{d}) joint Bayesian inversion in the case of fixed (\textbf{top}) and perturbed (\textbf{bottom}) acquisition geometry. The distributions are estimated using the kernel density estimate method from posterior samples \citep{parzen1962estimation}. Red lines denote the true velocity change.}
	\label{fig:marginals}
\end{figure}

For the separate Bayesian inversion strategy we follow the same procedure as described in the corresponding section above, except that different source locations and data are used in the monitoring inversion. The results are shown in Figure \ref{fig:separate_perturbedsource_results}. Overall the mean velocity change (Figure \ref{fig:separate_perturbedsource_results}c) shows similar features to that obtained in the case of fixed acquisition geometries. For example, the true velocity change can be observed clearly in the mean map and there are also many small scale random structures across area. Again the magnitude of standard deviation of velocity change is higher than that of velocity in the baseline inversion because of independence of the two inversions. A novel feature of these results is the set of dipping, slightly negative anomalies at depths $< 1$ km, which are therefore attributed to the perturbation in source locations.

For joint Bayesian inversion we conduct the inversion in the same way as above to invert for the baseline model and velocity change simultaneously. Overall the results show almost the same mean and standard deviation maps to those obtained in the case of fixed acquisition geometry for both the baseline velocity and velocity change (Figure \ref{fig:joint_perturbedsource_results}).

The above results show that the Bayesian methods are almost stable with respect to variations in the acquisition geometry, whereas the traditional double-difference algorithm is not. Furthermore, compared to the results obtained using separate Bayesian inversions, the results obtained using joint inversion show more accurate velocity changes and more reliable uncertainties because of the additional prior information imposed.

To further understand the results, in Figure \ref{fig:marginals} we show marginal distributions of velocity change obtained using the two Bayesian methods in the different cases along two vertical profiles whose locations are denoted by black dashed lines in Figure \ref{fig:separate_results}, \ref{fig:joint_results}, \ref{fig:separate_perturbedsource_results} and \ref{fig:joint_perturbedsource_results}. Similarly to above, the results obtained using separate Bayesian inversions show significantly broader distributions than those obtained using joint inversion because of the assumed independence between baseline and monitoring inversions in the former inversion strategy. For all results, the shallow part ($<$ 1.0 km) has lower uncertainty than deeper parts. Although the standard deviation models obtained using the joint inversion do not show lower uncertainty within the zone of velocity changes (Figure \ref{fig:joint_results}d and \ref{fig:joint_perturbedsource_results}d), the marginal distributions clearly reflect lower velocity within the area (Figure \ref{fig:marginals}b and d) which suggests that the velocity change is well constrained by the data and the prior information. By contrast, it is difficult to notice lower velocity from the distributions obtained using the separate inversion strategy because of their high uncertainty. Note that the marginal distributions obtained using the joint inversion show high probability density values at the boundaries of the prior distribution. This may be because the velocity change would have higher uncertainties if weaker prior information was imposed (i.e., the change is not well constrained by the data itself). As a result, when tight Uniform prior distributions are imposed, the mass of marginal distributions that would otherwise lie outside of the support of the Uniform distribution, concentrates close to its boundaries. It may also be possible that this is caused by biases of the algorithm itself, for example, the finite step we used in equation (\ref{eq:stochastic_svgd}); in practice the step length is always restricted by available computation power, and we have used the smallest size that was feasible. 

\subsection{Computational cost}
\begin{table}
	\begin{center}
		\begin{minipage}{160mm}
			\caption{A comparison of computational cost for the suite of inversion methods.} 	
			\begin{tabular}{ l@{\hspace{2cm}}l S[table-format=6.0]}
				\hline
				Method  & \multicolumn{2}{l}{Number of Simulations} \\
				\hline
				\multirow{2}{*}{Double difference inversion} &  baseline: & 163  \\[-5pt]
				                                             &  monitoring: & 50 \\
				\hdashline
				\multirow{2}{*}{Separate Bayesian inversion} &  baseline: & 6,000   \\[-5pt] 
				                                             & monitoring: & 2,000  \\
				\hdashline
				Joint Bayesian inversion & & 16,000 \\
				\hline		
			\end{tabular}		
			\label{tb:cost}
		\end{minipage}
	\end{center}
\end{table}
We summarize the number of simulations required by each method in Table \ref{tb:cost}. This provides a good metric of the overall computational cost because the forward and adjoint simulations are the most time-consuming components for each method. Note that because the inversions for fixed and perturbed acquisition geometries are conducted in the same way which results in the same number of simulations, we do not discriminate between the two cases in the table. Apparently the traditional double difference inversion is the most efficient method, but it cannot produce accurate uncertainty estimates, and it provides biased estimates when the acquisition geometry changes between surveys. The two Bayesian methods require significantly more computation than the double difference method. In addition, because in the joint inversion we simulate the baseline and monitoring data together at each iteration, the required number of simulations (16,000) is twice that required in the separate Bayesian inversion (8,000) even though the two inversions are conducted using the same number of iterations. However, the separate Bayesian inversion strategy does not provide useful uncertainty estimates for the velocity change due to the assumed independence of baseline and monitoring inversions. By contrast, the results obtained using joint inversion provide more accurate and useful uncertainty estimates because the method can take advantage of additional prior information on the velocity change itself. In addition, compared to the double difference inversion, both Bayesian methods provide stable and accurate mean velocity change estimates in the case of either fixed or perturbed acquisition geometries.     

Note that the above comparison depends on subjective assessments of convergence for each method, so the absolute computational time required by each method may not be entirely accurate. Nevertheless the comparison at least provides a reasonable insight into the efficiency of each method. To give an overall idea of the time required by the two Bayesian methods, the above inversions required 65 hours and 111 hours in wall time for the separate and joint inversions, respectively, both of which are parallelized using 40 AMD EPYC CPU cores.

\section{Discussion}

We demonstrated that Bayesian methods (separate Bayesian inversions of baseline and monitoring surveys, and a joint Bayesian inversion) can be used to detect velocity changes and quantify uncertainty for time-lapse inversions, and that they provide more accurate results than the traditional double difference method in the case where acquisition geometries were changed between the two surveys, even when the locations of sources and receivers were known exactly in each survey. This is because in double difference inversion the unexplained events of the baseline survey data are not compensated by the residual term $r(\mathbf{m}_{1})=\mathbf{u}(\mathbf{m}_{1})-\mathbf{d}_{1}$ in the new data $\mathbf{d}_{2}^{'}$ due to the change of source locations. As a result, those unexplained events can still affect the final time-lapse results. If the baseline model is perfect, that is, there are no unexplained events in the baseline survey data, the time-lapse change can be detected clearly even with perturbed acquisition geometries \citep{asnaashari2015time}. By contrast, Bayesian methods characterize the full Bayesian posterior distributions of seismic velocity in the baseline and monitoring surveys, or the full posterior distribution of velocity change between the two surveys. In either case, the obtained distribution contains information of time-lapse changes, regardless of whether there are perturbations in the acquisition geometries.

Bayesian methods are therefore particularly valuable when high repeatability of acquisition geometry is difficult to achieve or emulate by interpolation, for example, when source or receiver geometries are sparse. When using dense acquisition systems, time-lapse binning and data interpolation are usually applied to improve acquisition repeatability in standard double difference inversions, and the same can be applied in the Bayesian methods. In addition, the standard double difference inversion method demands an accurate baseline model, which may require more effort to build than in the Bayesian methods. For example, in the above study we inverted an extra low frequency data set in order to build an accurate baseline model, which may not always be available in practice. 

In this study we used a Uniform prior distribution on seismic velocities with a relatively large support (2 km/s), which leads to high uncertainty for velocity and consequently high uncertainty for velocity change. In practice where more knowledge about the subsurface is available, one can use a more informative prior distribution for the velocity. This will produce more accurate models and lower uncertainty for both velocity and velocity change. Note that when conducting separate Bayesian inversions the obtained uncertainty for velocity change is always larger than that obtained for velocity because of the implicit assumption of independence of baseline and monitoring inversions. In the joint inversion the velocity and velocity change are explicitly coupled, so strong prior information on velocity can also improve the accuracy of velocity change estimates. And of course, if an accurate baseline model is available and can be fixed during the inversion, one can also use the differential data between monitoring and baseline surveys and invert for the velocity change directly in the Bayesian inversion as in the standard double difference inversion \citep{kotsi2020uncertainty}.

We estimated the velocity change for the entire model area. This requires full model simulations during the inversion which can be computationally inefficient. If knowledge about locations of potential velocity change zones are available, one can also perform target oriented time-lapse inversions by assuming that the rest of the model is known. Then a local solver can be used to increase efficiency \citep{asnaashari2015time, kotsi2020uncertainty}. Alternatively, if it is not possible to perform target oriented inversion in practice, one can also use other faster, approximate forward modelling methods to improve efficiency, for example, neural network based modelling methods \citep{sirignano2018dgm, moseley2021finite}.

Although Bayesian inversion can produce more accurate results than the standard double difference inversion and can quantify uncertainty, it is also significantly more computational expensive. To improve efficiency of the methods, one might exploit high order gradient information, for example using a Hessian kernel function \citep{wang2019stein} or the stochastic Stein variational Newton method \citep{leviyev2022stochastic}. In addition, one can also use stochastic inversion by dividing the whole dataset into minibatches to reduce the computation cost as demonstrated by \cite{zhang20233}. 

The results obtained here may contain biases. For example, the small random structures in the velocity change model obtained using separate Bayesian inversions and those structures in the deeper part of the model obtained using joint inversion may constitute genuine biases due to lack of convergence of the algorithm. To further improve accuracy of the results, one may run the sSVGD algorithm longer. In addition, the discretization used in equation (\ref{eq:discretized_sde}) may cause errors and biases in results, a Metropolis-Hastings correction step can be added at each iteration \citep{metropolis1949monte, hastings1970monte}.

Although in this study we only applied Bayesian methods to 2D time-lapse change problems, the method should also be applicable to 3D cases since the sSVGD algorithm has already been used to solve 3D Bayesian FWI problems \citep{zhang20233}. However, because of the extremely high dimensional parameter space, it may not be easy for sSVGD to converge sufficiently, and consequently the time-lapse change may be difficult to obtain. In such cases stronger prior information on velocity might be required in order to detect velocity changes since this will reduce the computational complexity of converging to the solution. Alternatively one may try to reduce the dimensionality of the problem itself. For example, other parameterizations which use fewer parameters to represent the model may be used such as Voronoi tessellation \citep{bodin2009seismic, zhang20183}, Delaunay and Clough-Tocher parametrization \citep{curtis1997reconditioning}, wavelet parameterization \citep{hawkins2015geophysical}, discrete cosine transform \citep{kotsi2020time, urozayev2022reduced} and neural network parameterization \citep{laloy2017inversion, mosser2020stochastic, bloem2022introducing}. Other methods which project high dimensional spaces into lower dimension space may also be used to improve efficiency of the methods, for example, slice SVGD \citep{gong2020sliced} or projected SVGD \citep{chen2020projected}. 

\section{Conclusion}
In this study we explored two Bayesian inversion strategies: separate Bayesian inversions for baseline and monitoring surveys, and a joint Bayesian inversion of both survey data sets to solve time-lapse full waveform inversion (FWI) problems. We compared the results to those obtained using standard double difference inversion. The results show that all methods can provide accurate velocity change estimate in the case of fixed acquisition geometries, but in the case of perturbed acquisition geometries the two Bayesian methods produce significantly more accurate results than double difference inversion. In addition, Bayesian methods provide uncertainty estimates that account for the full nonlinearity of the model-data relationships, and any form of prior probability and data uncertainty distributions, which cannot be obtained using double difference inversion. However, when using the separate Bayesian inversion strategy the assumed independence between baseline and monitoring inversions causes the magnitude of the uncertainty estimate for velocity change to be higher than that for velocity itself, which makes the results less useful in practice. By contrast, the uncertainty estimates for velocity change from a single, joint Bayesian inversion are almost an order of magnitude smaller than those obtained from separate inversions because of additional prior information that can be imposed on the velocity change. This demonstrates that the joint inversion provides more accurate uncertainty estimates as the magnitude of velocity change is usually much smaller than that of velocity. We therefore conclude that Bayesian time-lapse FWI, especially joint Bayesian inversion, can be used to detect velocity change and to quantify associated uncertainties in time-lapse inversion and monitoring.  

\begin{acknowledgments}
The authors thank the Edinburgh Imaging Project sponsors (BP and Total) and National Natural Science Foundation of China (42204055) for supporting this research. This work has made use of the resources provided by the Edinburgh Compute and Data Facility (http://www.ecdf.ed.ac.uk/). 
\end{acknowledgments}

\bibliographystyle{gji}
\bibliography{bibliography}

\appendix

\label{lastpage}

\end{document}